\documentclass [12pt]{iopart}
\usepackage{lineno}
\begin{document}
\title[Residence time of energy in Earth's atmosphere and in the Sun]{Residence time of energy in Earth's atmosphere and in the Sun}
\author{C. Os\'acar$^1$, M.Membrado$^2$ and A.F. Pacheco$^3$}

\address{$^1$ Facultad de Ciencias. Universidad de Zaragoza. 50009 Zaragoza (Spain)}
\ead{cosacar@unizar.es}

\address{$^2$ Facultad de Ciencias. Universidad de Zaragoza. 50009 Zaragoza (Spain)}
\ead{membrado@unizar.es}

\address{$^3$ Facultad de Ciencias and BIFI. Universidad de Zaragoza. 50009 Zaragoza (Spain)}
\ead{amalio@unizar.es}

\begin{abstract}
In atmospheric chemistry, a parameter called {\it residence time} can be defined for each gas as $T=M/F$, where $M$ represents the average mass in the atmosphere and $F$ is the total average influx or outflux, which in time averages are equal. In this paper we extend this concept from matter to energy which is also a conservative quantity and estimate the average residence time of energy in Earth's atmosphere and in the Sun. For our atmosphere, the estimation amounts about 56 days and for the Sun the residence time is about $2.6 \times 10^7$ yr, which agrees with the Kelvin-Helmholtz time scale.
\end{abstract}

\par
\maketitle

\section{Introduction}
\label{intro}
When the inflow, $F_i$, of any substance into a box is equal to the outflow, $F_o$, out of the box, then the amount of that substance  in the box, $M$, is constant. This constitutes an equilibrium or steady state. Then the ratio of the stock in the box ($M$) to the flow rate (in or out $F_i=F_o=F$) is called the residence time and is a time scale for the transport of the substance in the box.
\begin{equation}
\label{tres}
t=\frac{M}{F}.
\end{equation}

We are refering  to a measurable and  conserved substance. A good example of this type is the parameter defined in atmospheric chemistry as the average residence time of each individual gas, defined as (\ref{tres}). $M$ is the total average mass of a gas in the atmosphere and $F$ the total average influx or outflux. In time the averages for the whole atmosphere $F_i$ and $F_o$ are equal. Another example used in climate studies is the amount of water in a reservoir\cite{reservoir}. In this case, $M$ is the amount of water stored in the reservoir and $F$ is the amount of water that goes in and out of the reservoir. \par
In this  paper we want to extend the substance that flows from matter to energy, and estimate the average residence time of energy in Earth's atmosphere and in the Sun. Obviously, in equation (\ref{tres}) $M$ and $F$ will now represent the total amount of energy in the system and the energy flux -in or out- respectively. Both cases correspond to steady state problems because the storage of energy in the atmosphere and in the Sun are not  systematically increasing or decreasing.\par
We want to emphasize that the ideas used in this paper are familiar in  undergraduate courses of atmospheric physics\cite{wallace} or astrophysics\cite{stellar}. \par
In section \ref{forms},  we consider the Earth's atmosphere as a big box and by using the appropriate energy data from \cite{peixoto} and \cite{climate} in Eq.(\ref{tres}), we compute the time of residence.
Due to the educational nature of this journal,  in section \ref{estim} we  compute the static energy of an air column using two academic methods. This will likely  increase the interest of this paper for the instructor  in undergraduate courses of atmospheric or environmental science\cite{cow}.  In the first method we perform a numerical integration of the pressure data offered in the US Standard Atmosphere. In the second method, we assume that pressure varies with height in an exponential way. \par
In section \ref{estimsun}, for the Sun, analogously to what has been done in section \ref{estim},  we have used Standard Solar Model tables to estimate the global Sun's energy. Dividing by the solar  luminosity, we estimate the residence time of energy in the Sun. This residence time agrees with  the Kelvin-Helmholtz time scale. In section \ref{discus} we present a brief discussion and our conclusions.

\section{Forms of energy in the atmosphere and time of residence.}
\label{forms}
The Earth's atmosphere is a comparatively thin film of gas distributed over the Earth's surface. This justifies the use of a planar atmosphere because the scale height for the vertical decrease in the air density is about 10 Km., which is very small compared with the planetary radius. For such a planar gaseous distribution, we will focus on the air contained in a vertical column with a base of surface unity.

In this section we use the energy data provided by Peixoto and Oort in \cite{peixoto}. The most important forms of energy in the atmosphere are: the thermodynamic internal energy, $U$, the potential energy due to Earth´s gravity, $P$, the kinetic energy, $K$, and the latent energy, $L$, related to the phase transitions of water.
\begin{equation}
E=U+P+K+L.
\end{equation}
The values quoted by these authors per unit surface in units of $10^7 \, \mathrm{J} \, \mathrm{m}^{-2}$ are:
\begin{equation}
U=180.3, \quad P=69.3, \quad K=0.123, \quad L=6.38, \quad E=256.1
\end{equation}
Note that  the sum $S=U+P= 249.6$, which constitutes the static energy of the atmosphere, is the main component. \par
For our purpose of computing the time of residence using (\ref{tres}), now we only need the inputs, or outputs, of energy in the atmosphere. For this purpose, we use the data cited by Schneider in Ref. \cite{climate}. It is common to express this data in units of $e=3.45 \, \mathrm{W} \, \mathrm{m}^{-2} $, because $e$ is a percentage of the solar irradiance out of the atmosphere. The atmosphere absorbs $25 \, e$ of solar energy, $29 \, e$ are absorbed  at the surface as sensible and latent heats, and finally it absorbs $100 \, e$ long wave radiation emitted by the surface. The radiation emitted by the surface is $104 \, e$, but $4 \, e$ escapes to space through the called atmospheric window. Thus the total energy input in the atmosphere is:
\begin{equation}
F_i=25 \, e +29 \, e +100 \, e= 154 \, e = 531 \, \mathrm{W} \, \mathrm{m}^{-2}.
\end{equation}
Regarding to the emitted energy flux, we identify two terms for the emitted energy flux, the component emitted towards space, $66 \, e$, and that emitted to the planet surface, commonly denoted as the greenhouse effect, $88 \, e$. The sum of the outgoing terms coincides with that of the ingoing terms, $F_o=F_i=F$. Thus, using the values of $E$ and $F$,  the estimation of the residence time of energy in our atmosphere, $t$, is:
\begin{equation}
t=\frac{E}{F}=\frac{256  \times 10^7 \mathrm{J} \, \mathrm{m}^{-2} } {531 \mathrm{W} \, \mathrm{m}^{-2}} = 4.82 \times 10^6 \, \mathrm{s} \approx 56 \, \mathrm{days}.
\end{equation}

\section{Estimation of the static energy in a column of air}
\label{estim}
In this section we will show a link between $U$ and $P$ and estimate the value of $S$ by means of two methods. For the planar atmosphere, the hydrostatic equation is 
\begin{equation}
\label{hydro}
- \rmd p =g \, \rho \, \rmd z.
\end{equation}
It states that the upward force due to the pressure gradient is equal to the weight of an infinitesimal layer of unit cross section. In (\ref{hydro}), $p$ is the pressure, $g$ the acceleration of gravity, $\rho$ the density and $z$ the height. \par
The potential energy of the air contained in the column mentioned in section \ref{forms}, is
\begin{equation}
\label{energy1}
P=\int_0^\infty g \, \rho \, z \, \rmd z.
\end{equation}
The upper limit $\infty$ stands for a height where pressure can be considered negligible.  Inserting (\ref{hydro}) into Eq.(\ref{energy1}), we obtain
\begin{equation}
\label{ocho2}
P= \int_{p_0}^0 -z \,\rmd p.
\end{equation}
And integrating by parts $\rmd (z \, p) = z \rmd p + p \, \rmd z$.
The result for $P$ is:
\begin{equation}
\label{nueve}
P=\int_{p_0}^0 -z \,\rmd p = \int_0^\infty \rmd (z\,p) +\int_0^\infty p \, \rmd z = 0 + \int_0^\infty p \, \rmd z.
\end{equation}
Regarding the thermal energy in the column, $U$, 
\begin{equation}
U= \int_0^\infty c_v \, \rho \, T \, \rmd z.
\end{equation}
Using the equation of state of ideal gases $p=\rho \, R \, T$ and inserting it in (\ref{ocho2})
\begin{equation}
P=\int_0^\infty p \, \rmd z = \int_0^\infty R \, \rho \, T \, \rmd z.
\end{equation}
Thus, as $c_v + R =c_p$
\begin{equation}
S=U+P =\int_0^\infty c_v \, \rho \, T \, \rmd z + R \int_0^\infty \rho \, T \rmd z = c_p \int_0^\infty \rho \, T \, \rmd z.
\end{equation}
$R$, $c_v$ and $c_p$ are the gas constant for dry air, and their specific heats at constant volume and constant pressure, respectively. As for ideal diatomic gases the ratio $c_p/R=3.5$, the computation of $S$ requires only the calculation of the pressure integral, $P$.
\begin{equation}
S= 3.5 \, R \int_0^\infty T \, \rho \, \rmd z = 3.5 P.
\end{equation}
\par
A first method (subindex 1) to estimate $P$ is using the pressure data provided by the US Standard Atmosphere \cite{USStandAtm}. We have approximated the integral (\ref{nueve}) as the sum:
\begin{equation}
P_1=\sum_0^{80 \, \mathrm{Km}} p_i \Delta z_i.
\end{equation}
The results for $P_1$ and $S_1$ are:
\begin{eqnarray}
P_1= 7.55 \times 10^8 \, \mathrm{J \, m}^{-2} \\
S_1= 2.64 \times 10^9 \, \mathrm{J \, m}^{-2}.
\end{eqnarray}
An alternative second method (subindex 2)  to compute $P$ is to assume that  pressure varies with height in an exponential  way:
\begin{equation}
\label{expon}
p(z)=p_0 \exp(-z/H).
\end{equation}
We know from the hypsometric equation that pressure is rigorously exponential with height only if the atmosphere is isothermal. Notwithstanding, using this exponential hypothesis with $p_0=1.01325 \times 10^5$ Pa, and a height scale $H \approx 6.5$ Km, the fit of the global data pressure (not only in the troposphere) is sufficiently good for our purposes. 
Thus, adopting (\ref{expon}), the results for $P$ and $E$ are:
\begin{eqnarray}
P_2=p_0 \, H = 6.58 \times 10^8 \, \mathrm{J \, m}^{-2} \\
S_2= 3.5 p_0 \, H = 2.3 \times 10^9 \, \mathrm{J \, m}^{-2}
\end{eqnarray}
Note that both approximate methods provide the correct order of magnitude for the energy of a column of air but the first one overestimates $S_1$  with respect to $E$.
 
\section{Estimation of the solar energy and the time or residence of energy in the Sun}
\label{estimsun}
We assume that a star is in hydrostatic equilibrium under its own gravitation, i.e. the inward directed gravitational force exactly balances the outward directed force due to the gas and radiation pressure. The equation of hydrostatic equilibrium reads as follows:
\begin{equation}
\frac{\rmd p}{\rmd r}= -\rho \frac{\rmd \phi}{\rmd r}.
\end{equation}
Where $\phi$ is the gravitational potential of a spherical mass distribution and denoting by $M_r$ the mass contained in a sphere of radius $r$
\begin{equation}
\frac{\rmd \phi}{\rmd r} = \frac{G M_r}{r^2}.
\end{equation}
Combining the two previous equations, we obtain
\begin{equation}
\label{dp}
\rmd p = -\rho \frac{G M_r}{r^2} \rmd r
\end{equation}
Now, we multiply both sides of (\ref{dp}) by $4 \pi r^3$ to obtain
\begin{equation}
4 \pi r^3 \rmd p = -4 \pi r^3 \rho \frac{G M_r}{r 2} \rmd r=-4 \pi \rho G M_r r \rmd r,
\end{equation}
and integrate between $r=0$ and $r=R$
\begin{equation}
\label{egrav1}
\int_{p(r=0)}^{p(r=R)} 4 \pi r^3 \rmd p = \int_0^R -4 \pi \rho GM_r r \rmd r =E_{grav}.
\end{equation}
The integral of the left can be performed by parts. Denoting
\begin{equation}
u=r^3, \quad v=p, \qquad \rmd (r^3 p)=3 r^2 \rmd r p+r^3 \rmd p.
\end{equation}
Inserting this information in (\ref{egrav1})
\begin{equation}
E_{grav}=-3 \int_0^R p \, \rmd r^3=-3 {\mathcal P}
\end{equation}
where ${\mathcal P}$ stands for
\begin{equation}
\label{egrav}
{\mathcal P}=\int_0^R p \rmd r^3.
\end{equation}
For bound states, like the Sun, where the attractive force is of the type $r^{-2}$, the Virial Theorem states that:
\begin{equation}
E_{therm}= -E_{grav}/2.
\end{equation}
And therefore, the total energy fulfills
\begin{equation}
E=E_{grav}+E_{therm}=\frac{-3}{2} {\mathcal P}
\end{equation}
Note that the total energy, $E$, the gravitational energy, the thermal energy and the integral ${\mathcal P}$ are all of the same magnitude.
The order of magnitude of the solar energy, $E$, can be estimated using the pressure data of a Standard Solar Model of the Sun as those found, for example, in \cite{stixbook}. \par
Using that data, we have have approximated eq.(\ref{egrav}) by the sum of products of the volume of a shell comprised between $r_i$ and $r_{i+1}$ times the average pressure in that shell:
\begin{equation}
\label{solsum}
{\mathcal P} = \int_0^{R_\odot} p(r) \, 4 \pi \, r^2 \, \rmd r  \approx \sum_0^{R_\odot} \frac{4 \pi}{3} \left(r_{i+1}^3 -r_i^3 \right) \left ( \frac{p_{i+1}+p_i}{2} \right).
\end{equation}
Where the index $i$ runs from the center up to the surface of the Sun. The result of the sum (\ref{solsum}) times 3/2, is the estimation of $ \| E \| \approx 10^{41}$ J.

The ratio between $ \| E \|$ and  the solar luminosity,  $L_\odot = 3.82 \times 10^{26} \, \mathrm{W}$, is our result  for the energy residence time in the Sun:
\begin{equation}
\label{tsol}
t_\odot \approx 2.6 \times 10^7 \mathrm {yr}.
\end{equation}
The Kelvin-Helmholtz  (K-H) time scale, see for example \cite{stellar}, for the Sun is:
\begin{equation}
t_{KH}=\frac{G M_\odot^2}{R_\odot L_\odot}=3  \times 10^7 \, \mathrm{yr}.
\end{equation}
This time scale is roughly the time needed by the star to settle to equilibrium after a global thermal perturbation. \par
As mentioned above, for stars in equilibrium, the  total energy and  the gravitational energy are of the same
order of magnitude .  The K-H time scale is nothing but 
\begin{equation}
t_{KH} \approx \frac{\| E_{grav} \|}{L_\odot}.
\end{equation}
Therefore the time of residence of energy computed for the Sun  in (\ref{tsol}) is basically the same thing as the classical  K-H  time scale. Stix also estimated the same order of magnitude in \cite{stixpaper} for the time scale of energy transport in the Sun.

\section{Discussion}
\label{discus}

In this paper, we have considered our atmosphere as a big box where energy is in equilibrium,  and have estimated its residence time. It amounts to about  56 days.  When the same idea is applied to the Sun, we obtain $t \approx 2.6 \times 10^7 \, \mathrm{yr}$. \par

In astrophysics, the question: “how long a photon might take to get from the core of the Sun to the surface” has frequently been put forward. The answer of several authors was $\approx 10^4 \, \mathrm{yr}$,  see \cite{shu, bahcal}, etc. In 1989, Mitalas and Sills \cite{mitalas} pointed out that the average step length assumed by previous authors for a photon diffusing through the Sun was too long. Correcting this step length, they obtained $1.7 \times 10^5 \, \mathrm{yr}$.  Finally, invoking  the large heat capacity of the interior of the star, Stix \cite{stixpaper} corrected the previous result up to a time scale of the order of $3 \times 10^7 \, \mathrm{yr}$, i.e. the K-H time scale. \par
Bearing in mind the explanation given in section \ref{estimsun}, our conclusion for the residence time of  energy in Earth's atmosphere ($t \approx 56 \, \mathrm{days}$) is that this is the equivalent of the K-H time scale for the Sun. Thus, after a global thermal perturbation, the atmosphere would need about a couple of months to return to equilibrium. \par

The strategy of computing the total energies by the simple way of using tables, US Standard Atmosphere and a Solar Standard Model that can be found in internet, is in our opinion, a very instructive task for a class of undergraduate students.

\end{document}